\documentclass[superscriptaddress,aps,twocolumn]{revtex4-1}
\usepackage{bm}
\usepackage{float}
\usepackage{amssymb}
\usepackage{amsmath}
\usepackage{multirow}
\usepackage{lineno}
\usepackage{graphicx}
\usepackage{comment}
\usepackage{soul}
\usepackage[colorlinks=true,citecolor=blue,linkcolor=blue]{hyperref}
\usepackage[usenames]{color}
\usepackage{subfigure}

\begin{document}

\title{Generation of isolated attosecond electron bunches by the diffraction of a polarization-tailored intense laser beam}

\author{Ke Hu}
\affiliation{Tsung-Dao Lee Institute, Shanghai Jiao Tong University, Shanghai 200240, China}
\affiliation{Collaborative Innovation Center of IFSA (CICIFSA), Shanghai Jiao Tong University, Shanghai 200240, China}
\author{Longqing Yi}
\thanks{lqyi@sjtu.edu.cn}
\affiliation{Tsung-Dao Lee Institute, Shanghai Jiao Tong University, Shanghai 200240, China}
\affiliation{Collaborative Innovation Center of IFSA (CICIFSA), Shanghai Jiao Tong University, Shanghai 200240, China}
\affiliation{Key Laboratory for Laser Plasmas (Ministry of Education), School of Physics and Astronomy, Shanghai Jiao Tong University, Shanghai 200240, China}

\date{\today}

\begin{abstract}
We propose utilizing a polarization-tailored high-power laser pulse to extract and accelerate electrons from the edge of a solid foil target to produce isolated attosecond electron bunches. The laser pulse consists of two orthogonally-polarized components with a time delay comparable to the pulse duration, such that the polarization in the middle of the pulse rapidly rotates over 90$^\circ$ within few optical cycles.
Three-dimensional (3D) Particle-in-Cell simulations show that when such a light pulse diffracts at the edge of a plasma foil, a series of isolated relativistic electron bunches are emitted into separated azimuthal angles determined by the varying polarization.
In comparison with most other methods that require an ultra-short drive laser, we show the proposed scheme works well with typical multi-cycle ($\sim 30~$fs) pulses from high-power laser facilities.
The generated electron bunches have typical durations of a few hundred attoseconds and charges of tens of picocoulombs.
\end{abstract}
\maketitle

The production of attosecond electron bunches has become a research focus
because of its diverse applications, such as
ultrafast electron imaging \cite{Baum2009,Shorokhov2016},
electron diffraction and microscopy \cite{Priebe2017,Hassan2018,Morimoto2018}.
In particular, such beams can serve as secondary sources for radiation production down to attosecond-level \cite{Luo2013,Zhu2018,Wu2010,Wu2011,Dromey2012}.
There have been extensive studies on generating relativistic attosecond electron bunches using nonlinear interaction between an intense femtosecond laser pulse and matter, both theoretical  \cite{Naumova2004,Popov2009,Liseykina2010} and experimental \cite{Sears2008,Morimoto2018}.
However, the generated electron beam typically consists of a train of bunches.
The production of an Isolated Attosecond Electron Bunch (IAEB) remains challenging.
To the best of our knowledge, no experimental evidence of IAEBs has been reported so far.
Moreover, most current numerical approaches rely on interactions between a few-cycle driving laser pulse and a nano-scaled target, such as a droplet \cite{Horny2021}, a nanotip \cite{Cardenas2019}, or an ultrathin foil \cite{Kulagin2007,Wu2011}.
The drive laser typically has a Full Width at Half Maximum (FWHM) duration smaller than 5 fs, ensuring that only one optical cycle is sufficiently strong to extract a considerably populated bunch from the target.

In this letter, we demonstrate that the generation of IAEBs can be achieved by the diffraction of an intense polarization-tailored laser pulse at the edge of a solid foil.
The drive laser is multi-cycle, which is available in most high-power facilities. It is linearly polarized in orthogonal directions at both ends, and in the middle the polarization direction rapidly rotates over 90$^\circ$ within few optical cycles (Fig.1(a)).
Such pulses can be routinely produced by the polarization gating technique \cite{Corkum1994,Tcherbakoff2003,Sola2006},
where a quartz plate is used to split a linearly polarized laser pulse into two orthogonally-polarized components, with an adjustable delay by varying the thickness of the plate.
When such a laser shines at the overdense plasma edge, relativistic electrons are accelerated via vacuum laser acceleration (VLA) \cite{Malka1997,Thevenet2016}, with the emission angles controlled by the laser polarization.
Thus, attosecond electron bunches are produced in each laser cycle from the middle of the pulse, but emitted in slightly different angles,
analogous to the ``attosecond lighthouse" \cite{Vincenti2012,Wheeler2012}, 
except that the emission angle is controlled by a rotating polarization instead of the wavefront direction.

\begin{figure}[t]
\centering
\includegraphics[width=0.48\textwidth]{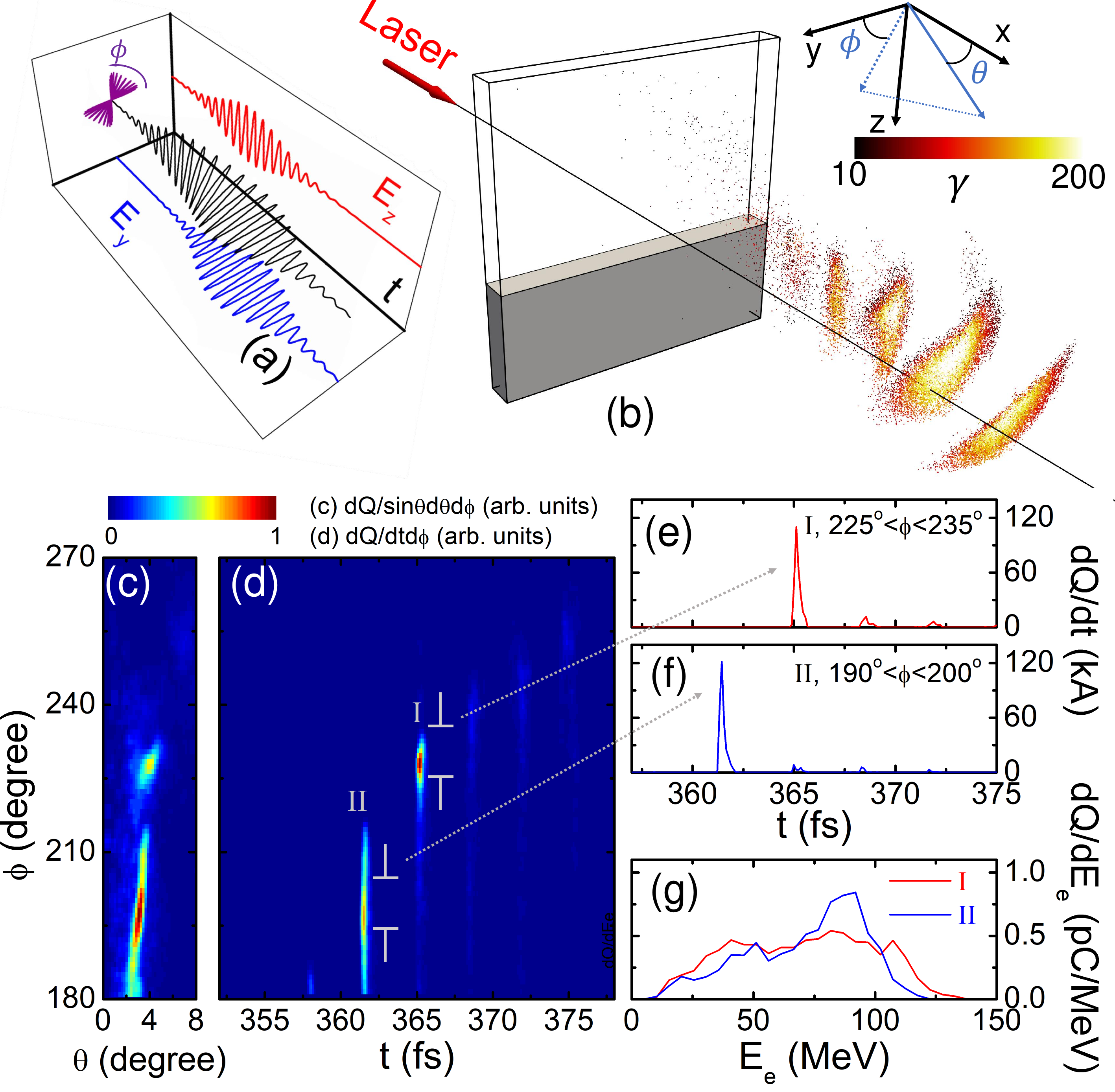}
\caption{
(a) Electric field of the polarization-tailored drive laser in $E_y-E_z-t$ space and its projections on $E_y-t$, $E_z-t$ and $E_z-E_y$ planes.
(b) 3D schematic setup of the proposed scheme. The orange dots are fast electrons ($\gamma>10$) at simulation time $t=120\tau_0$ and the color represents their energies. Density distribution of fast electrons are shown in $\phi-\theta$ space (c) and $\phi-t$ space (d).
(e)(f) Currents of the Bunch I and Bunch II, respectively. The two bunches are obtained by selecting out electrons within a certain range of the azimuthal angle (gray area in (d)).
(g) Energy spectra of Bunch I and II.
}
\end{figure}

The sketch of our simulation setup is illustrated in Fig. 1(b), the simulation is performed with three-dimensional (3D)
particle-in-cell (PIC) simulations with the EPOCH code \cite{Arber2015}.
A polarization-tailored drive laser pulse travelling along the $x$-direction  is focused onto the edge of a solid foil target.
The laser focus spot is partially blocked by the foil, which leads to diffraction.
The laser field is given as
\begin{equation}
\begin{aligned}
\mathbf{E}&=
\text{exp}[-(y^2+z^2)^2/w_0^2]\\
\cdot
&[\mathbf{e_y}E_1\text{exp}(-t^2/T^2)\text{exp}(ik_0x-i\omega_0t)\\
&+\mathbf{e_z}E_2\text{exp}(-(t-\Delta t)^2/T^2)\text{exp}(ik_0x-i\omega_0t+\varphi)].
\end{aligned}
\end{equation}
One can see that the laser is consisted of two orthogonally polarized components with amplitudes $E_1$ and $E_2$, denoted by unit vectors $\mathbf{e_y}$ and $\mathbf{e_z}$.
Here we first consider the case with $E_1 = E_2$, and the normalized amplitudes are $a_1 = a_2 = eE_1/mc\omega_0 = 10$,
where $\omega_0$ is the angular frequency, $c$ is the speed of light, $m$ is the electron mass, and $e$ is the unit charge.
The laser wavelength is $\lambda_0=1\ {\mu \rm m}$, $k_0=2\pi/\lambda_0$ is the wavenumber, $\tau_0=\lambda_0/c$ is the laser period, and $w_0=12\ {\mu \rm m}$ is the laser spot size.
The two laser components, each has a FWHM duration of $20.0\ {\rm fs}$ ($T=17.0\ {\rm fs}$) are separated by a time delay of $\Delta t=23.3$ fs.
The relative phase difference is set to be $\varphi = 0$ here, its effects will be discussed in the remainder of this work.

The foil target is modeled by a $1$-${\mu \rm m}$ thick (in the $x$ direction) pre-ionized plasma slab, 
and the density is $n_0=50n_c$, where $n_c=\epsilon_0m\omega_0^2/e^2$ is the critical density.
The foil is place at $z>z_0=4.4\ {\mu \rm m}$, with
a density ramp at the boundary to account for plasma expansion due to the laser prepulse: $n(z)=n_0{\rm exp}((z-z_0)/\sigma_0)$ for $z<z_0$, where the sale length is $\sigma_0=0.1\ {\mu \rm m}$.
The simulation box has dimensions of $x\times y \times z=32\times30\times30\ \mu{\rm m}^3$ and is sampled by $1280\times 1200\times 1200$ cells, with sixteen macroparticles for electrons and two for C$^{6+}$ per cell.
A moving window is used to improve computational efficiency.

When there is a sufficiently strong laser electric field component perpendicular to the plasma-vacuum boundary, the diffraction of a relativistic laser can produce a train of fast electron bunches \cite{Naumova2004}.
In our simulation, the edge of the target is parallel to the laser polarization in the leading half.
Then the laser polarization rotates over 90$^{\circ}$ to the perpendicular direction ($E_z$), thus electrons can be extracted out of the foil and accelerated.

The electron bunches, recorded at $100\ {\mu \rm m}$ away from the target are shown as orange dots in Fig.~1(b).
Their distributions in $\phi-\theta$ space (divergence), and $\phi-t$ space are illustrated in Figs.~1(c-d), where $\theta$ and $\phi$ are defined in Fig.~1(b). Notably, all the electron bunches are forward directed, concentrated within a small opening angle $d\theta\approx 1^\circ$ centered around $\theta\approx 3^\circ$.
On the other hand, the azimuthal angles of the accelerated electrons are strongly dependent on the instantaneous polarization, which is rapidly rotating.
As shown by Fig.~1(d), two main electron bunches are generated with an azimuthal angle separation of $\Delta\phi_s\approx30^\circ$.
Notably, this separation exceeds the azimuthal divergence observed within each individual bunch ($\Delta\phi_b\approx8^{\circ}\ \&\ 20^{\circ}$ for the two bunches).
Therefore, IAEBs can be obtained by collecting the electrons within certain ranges of the azimuthal angle.

Electrons that peak around $\phi = 235^\circ$ and $195^\circ$ are selected out and denoted as Bunch I and Bunch II, and the beam current of electrons centered within $10^\circ$ around each peak are plotted in Figs. 1(e-f).
The temporal FWHM duration of Bunch I and II are 250 as and 200 as, with total charges of $32.0$ pC and $33.7$ pC, respectively. This results in an averaged current of 127.88 kA for Bunch I and 168.3 kA for Bunch II.
Fig. 1(g) shows the broad energy spectra of the two bunches, as can be expected from the VLA mechanism.
The maximum energies are $120.0$ MeV for Bunch I and $130.9$ MeV for Bunch II, the averaged electron energy are around $67$ MeV for both bunches. Thus, the mean current of the IAEB reach $\sim10\%$ of the Alfv\'en limit.

To produce such an ultrashort dense electron bunch, it is crucial to accelerate the electrons to relativistic energy as fast as possible, so that the longitudinal expelling force between them are reduced by $\gamma^{-2}$ \cite{Jackson1998}, and the electrons in the bunch quickly become heavy before they can expand.

In the present scheme, the diffraction of the drive laser beam can boost the electron energy significantly, as shown by Fig.~2(a).
By tracking the trajectories of electrons in Bunch I, we can see that a brief but strong acceleration phase boosts the electron energy to $\gamma\sim10$ during laser diffraction.
Then, the electrons are accelerated mainly by the VLA mechanism in a second, much longer stage, with moderate acceleration gradient.

In the first stage, the main contribution comes from the strong longitudinal component $E_x$ of the near-field diffracted light (Fig. 2(b)).
The electrons are injected into the accelerating phase, with velocities predominately along the $x$ direction, thus experience an acceleration gradient comparable to the drive laser field, which is roughly 3 times higher than that in the VLA stage.
Then, a short deceleration occurs, as the electrons cannot keep up with the accelerating phase.
Nevertheless, since the diffracted laser beam disperses very quickly, the $E_x$ field is much weaker when dephasing occurs.
Therefore, a net energy gain can be expected.
When the electrons are sufficiently far away from the target, the electromagnetic field is essentially a planar wave, so that VLA takes over.

\begin{figure}[t]
\centering
\includegraphics[width=0.48\textwidth]{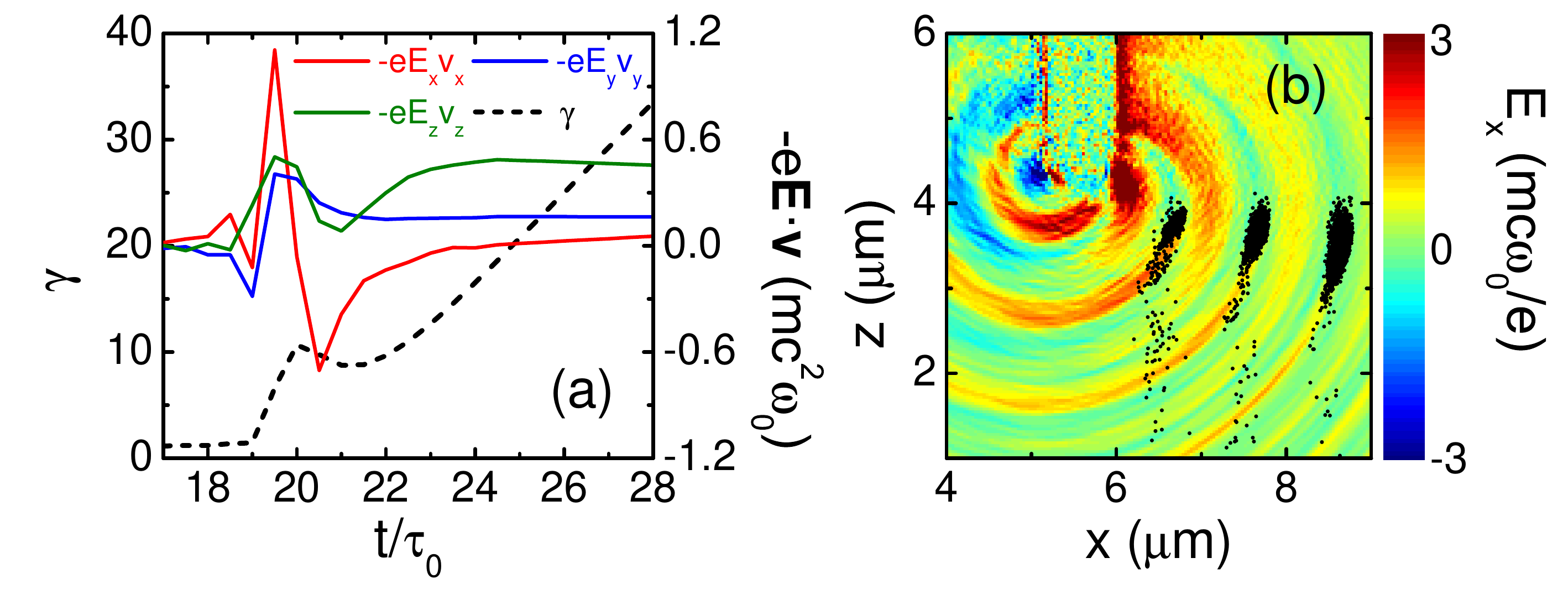}
\caption{
Electron energy boost by the diffracted laser field.
(a) Average $\gamma$ factor and the instantenous work done by each electric field components plotted against time at the early stage of acceleration.
(b) Locations of fast electrons ($\gamma>5$) on the $x-z$ plane at $t=21\tau_0$. On the background is a snapshot of $E_x$ at the same moment.
}
\end{figure}

In the following, we consider the condition under which the IAEB can be produced. Namely, the divergence of each electron bunch should be smaller than the bunch separation angle, $\Delta\phi_b<\Delta \phi_s \approx \pi/2N_{ol}$, where $N_{ol}\approx (2T-\Delta t)/\tau_0$ is the number of overlapping optical cycles.

The VLA mechanism follows the conservation of canonical momentum in transverse direction $\mathbf{p}_\perp = e\mathbf{A}/c$ \cite{Gibbon2005}, where $\mathbf{A}$ is the vector potential satisfying $\mathbf{A}=-\partial\mathbf{E}/\partial t$.
The azimuthal angle of the electrons is determined by the electric fields $\phi_b(t)=\arctan(E_z(t)/E_y(t))$, since $E_z/E_y=A_z/A_y$.
Apparently, this value varies in each optical cycle in the polarization-rotating region of the laser.  As a result, the electrons in Bunch I and II [Fig.~1(d)] form two distinguished peaks  in transverse momentum space $p_y - p_z$, marked by black and red dots in Fig.~3(a), respectively, which can be fitted by the ratio of instantaneous electric fields act on them.

The azimuthal divergence of each bunch $\Delta\phi_b$ can be obtained by considering the variation of $E_y$ and $E_z$ components within the bunch duration $\delta_b$. This is predominately determined by $\delta_b$ and the relative phase difference ($\varphi$) between $E_y$ and $E_z$ components.
The electrons are injected within a narrow phase close to the peak of the laser electric field perpendicular to the diffraction edge ($E_z$) in each optical cycle,  this can be seen from Fig.~3(b), where the relative phase between the $E_y$ and $E_z$ fields is $\varphi = 0.15\pi$.
The divergence of the electron bunches can be obtained as $\Delta\phi_b\approx  \arctan(1/\cos(\varphi+\varphi_\delta)) - \arctan(\cos\varphi_\delta)$, where $\varphi_\delta = 2\pi\delta_b/\tau_0$ is corresponding to the accelerating phase occupied by the electron bunch.
Here we have approximately set the ratio of local $E_z/E_y$ to be unity, which can be justified by our PIC simulations. 

 \begin{figure}[t]
\centering
\includegraphics[width=0.48\textwidth]{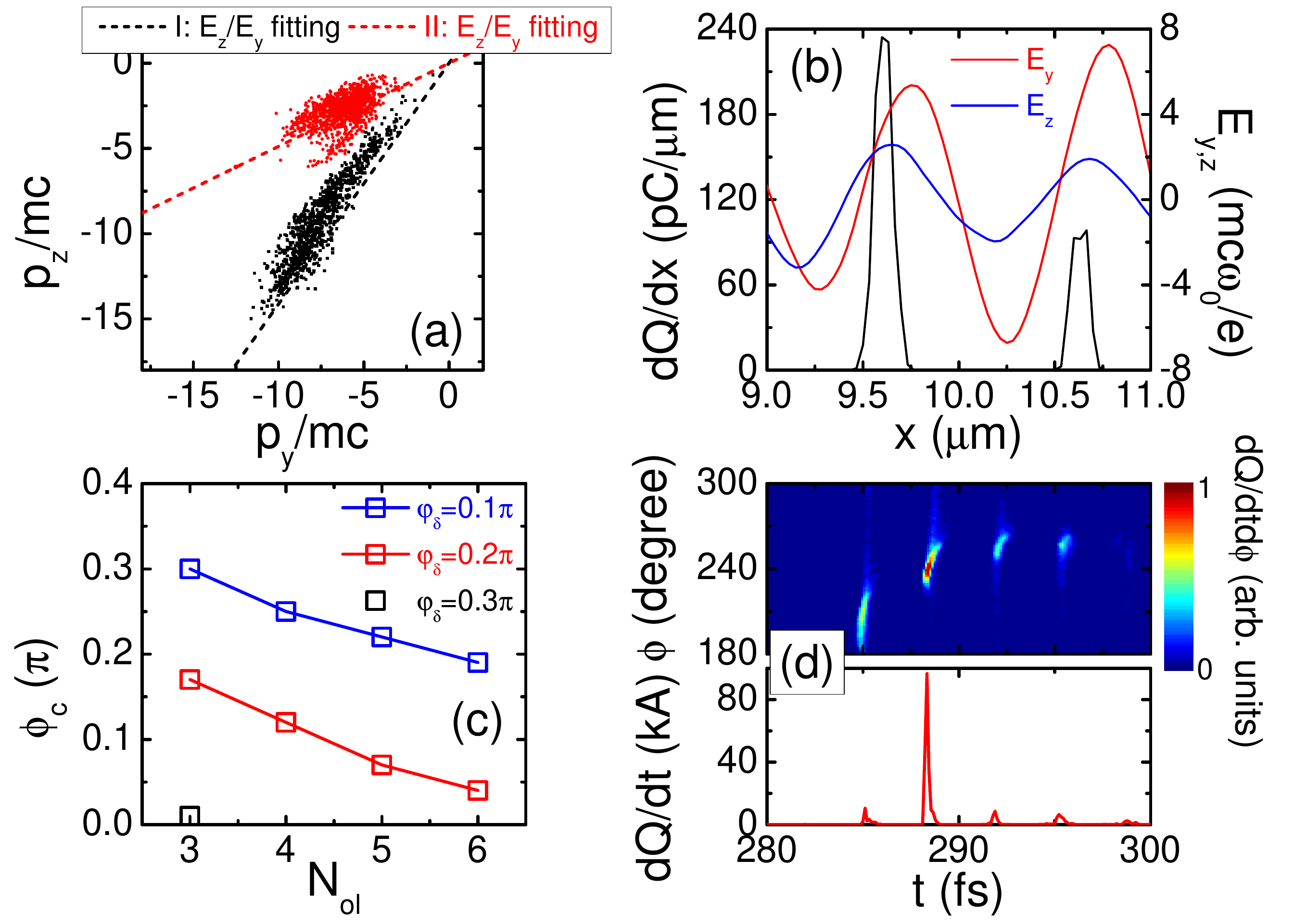}
\caption{
(a) Electron distributions of Bunch I and II in $p_y-p_z$ space at $t=80\tau_0$. Each dashed line represents a linear fit of the ratio of the electric field acting on the corresponding electrons
in the $z$ and $y$ directions.
(b) Waveforms of $E_y$ and $E_z$, and density profile of the fast electrons at $t=22\tau_0$.
The relative phase is $\varphi=0.15\pi$.
(c) Critical phase mismatch $\phi_c$ versus the number of overlapping optical cycles $N_{ol}$ under different accelerating phases occupied by electrons $\varphi_{\delta}$.
(d) Simulation results of the case with $T=4\tau_0$, $\Delta t=5\tau_0$, and $\varphi=0.2\pi$.
Upper: density distribution of fast electrons in $\phi-t$ space.
Lower: current of the IAEB (electrons within $235^{\circ}<\phi<245^{\circ}$).
}
\end{figure}

Due to the ultrashort nature of attosecond bunches, $\varphi_\delta$ is typically quite small ($<0.2\pi$).
It is easy to see that $\Delta\phi_b$ is in minimum when $\varphi = 0$ and increases with $\varphi$.
In reality, it is very challenging to manipulate the relative phase between $E_y$ and $E_z$, as it requires sub-cycle level temporal precision.

Fortunately, we can demonstrate that by controlling the number of overlapping optical cycles $N_{ol}$, our scheme works with a substantial range of non-zero $\phi$. The results are summarized in Fig.~3(c), where the blue, red, and black curves show the evolution of critical phase mismatch $\varphi_c$, below which the IAEBs can be produced, as an function of $N_{ol}$ for electron bunch length of $166.7~\rm as$, $333.3~\rm as$, and $500~\rm as$, respectively.
Here we only consider $0\leq\varphi<\pi/2$, for $\pi/2\leq\varphi<\pi$, the results are the same, except the IAEBs are emitted into a different quadrant in the $y$-$z$ plane.

As one can see, it is in general more resilient against wider ranges of phase mismatch for shorter electron bunches. Importantly, we show by reducing $N_{ol}$, $\varphi_c$ increases significantly,
meaning a higher tolerance for the phase difference.
According to our simulations, $\delta_b$ is typically 100-300 as, corresponding to $\varphi_c\sim0.2\pi - 0.3\pi$. An example is presented in Fig.~3(d), where two 15.7-fs mutual orthogonally-polarized pulses are separated by $\Delta t = 5\tau_0$ ($N_{ol}=3$), an isolated 200-as electron bunch is produced with $\varphi = 0.2\pi$.

It is important to note that the number of overlapping optical cycles ($N_{ol}$) cannot be arbitrarily small for a given laser duration.
In particular when the duration is long, a small $N_{ol}$ may cause the intensity of the laser beam in the polarization-tailored region too weak to extract electrons out of the plasma target. In this work, we restrict ourselves to laser durations $T>15 $ fs and $N_{ol}>3$.

\begin{figure}[t]
\centering
\includegraphics[width=0.48\textwidth]{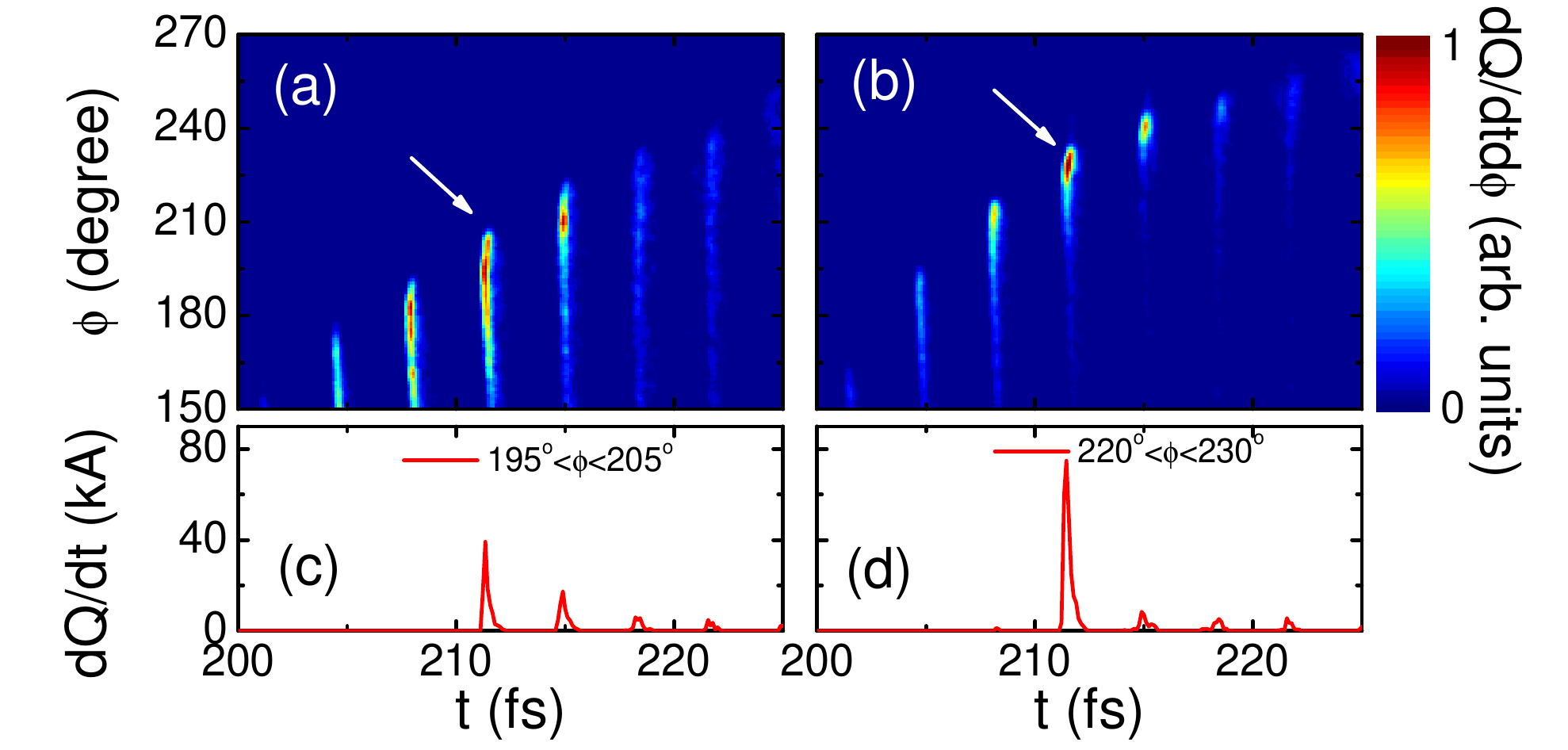}
\caption{Results of 3D PIC simulations investigating the preheating effect.
 Density distributions of fast electrons for $a_1=5$, $a_2=10$ (a) and $a_1=a_2=10$ (b) are plotted in $\phi-t$ space.
 (c)(d) Currents of the IAEBS, selected from the electron bunches indicated by the arrows in (a) and (b), respectively.
 The laser FWHM duration is $27.5\ {\rm fs}$ for each component, and the two components are separated by $33.3\ {\rm fs}$.
}
\end{figure}

Finally, we consider the preheating effect which leads to a non-negligible initial electron momenta ($\mathbf{p_{\perp,0}}$) before they are pulled out by the laser.
In this study, the preheating is mostly caused by the first half of the main laser pulse, which leads to ``$\mathbf{J}\times \mathbf{B}$" heating \cite{Kruer1985} or vacuum heating \cite{Brunel1987} of the target electrons. This effect becomes increasingly important as the laser duration becomes longer. Here we neglect the heating by the laser prepulse as the energy in the prepulse is much lower than the main pulse for a reasonably high-contrast laser beam.

The VLA stage satisfies $\mathbf{p_\perp} = \mathbf{p_{\perp,0}}+e\mathbf{A}/c$ when considering the
initial momenta.
As discussed before, the second term ($e\mathbf{A}/c$) is quite similar for all the electrons in the same bunch, since they occupy a small phase of the accelerating laser beam, but a boarder range of $\mathbf{p_{\perp,0}}$ distribution can significantly increases the azimuthal angular divergence of the bunches, making it harder to separate them from each other.
As shown in Fig.~4(a), where we consider a longer laser pulse with $T=7\tau_0$ and $\Delta t=10\tau_0$ ($N_{ol}\approx 4$) (other parameters are the same as in Fig.~1), the angular divergence is $\Delta\phi_b\sim30^\circ$.

Nevertheless, the preheating effect can be mitigated by reducing the intensity of the leading-half component, as shown by Fig.~4(b).
When we set $a_1=5, \ a_2=10$, one can see the azimuthal beam divergence is significantly reduced ($\Delta\phi_b\sim10^\circ$) comparing to Fig.~4(a).
This can be easily understood as the leading half of the drive laser beam is weaker, thus the plasma temperature is lower, the accelerated electrons tends to follow the same trajectory.  Therefore, the intensity ratio between two laser components $a_2/a_1$ provides an additional degree of freedom to control the electron emission.
In the case presented in Fig. 4(b), the electron charge collected within
$220^{\circ}<\phi<230^{\circ}$ is $27.8\ {\rm pC}$, with a temporal FHWM duration of 300 as, corresponding to an averaged beam current of 92.7 kA, which is comparable to the short duration case presented in Fig.~1.

In summary, our study demonstrates the generation of isolated attosecond electron bunches with charge at 10-pC-level when a polarization-tailored femtosecond laser beam diffracts at the edge of a solid foil.
The extracted electrons are first boosted to relativistic energies by the near-field diffracted laser field, and then accelerated by the laser fields with tailored polarization, emitting into separated azimuthal angles.
The number of overlapping optical cycles between two orthogonally polarized components, as well as their intensity ratio provide two major degrees of freedom to control the electron emission process.
The proposed scheme can work with a much longer laser duration compared with most other laser-plasma sources of isolated attosecond electron bunches.

\begin{acknowledgments}
This work is supported by the National Key R$\&$D Program of China (No. 2021YFA1601700), and the National Natural Science Foundation of China (No. 12205187).
\end{acknowledgments}



\end{document}